\documentclass[aps,prl,nofootinbib,showpacs,twocolumn]{revtex4-1}
\usepackage{amssymb,latexsym,mathrsfs,color}
\usepackage{dcolumn}
\usepackage{bm}
\usepackage{amsfonts}
\usepackage{amsmath}
\usepackage{graphicx}

\newcommand{\beq}{\begin{equation}}
\newcommand{\eeq}{\end{equation}}
\newcommand{\beqn}{\begin{eqnarray}}
\newcommand{\eeqn}{\end{eqnarray}}
\newcommand{\bearr}{\begin{array}}
\newcommand{\enarr}{\end{array}}

\def\bea{\begin{eqnarray}}
\def\eea{\end{eqnarray}}
\def\ba{\begin{array}}
\def\ea{\end{array}}
\def\n{\nonumber}
\def\c{\mathscr}

\def\c{\mathscr}
\def\pre{Phys. Rev. E}

\begin{document}

\title{Absorbing Phase Transition in Energy Exchange Models} 
\author{Urna Basu, Mahashweta Basu and P. K. Mohanty}
\affiliation{TCMP Division, Saha Institute of
Nuclear Physics, 1/AF Bidhan Nagar, Kolkata 700064, India.}

\begin{abstract}
We  study energy exchange models  with dissipation ($\lambda$)  and noise  (of amplitude $\sigma$)  
and show that in presence of a threshold  these models  undergo an absorbing phase transition when  either of 
dissipation  or noise  strength or both are varied. Using Monte Carlo simulations  we find that 
the  behaviour along the  critical line, which separates the active  phase  from the absorbing  one,    
belongs to Directed Percolation (DP) universality class. We  claim that the  conserved version with $\lambda=1$ and $\sigma=0$  also shows a DP transition; the apparent  non-DP behaviour observed  earlier is an artifact of undershooting in the decay of activity density  starting from a 
random initial condition. 
\end{abstract}

\maketitle

\section{Introduction}

Absorbing state phase transitions (APT)  \cite{DPbook,book} refer to a special class of non-equilibrium phase transitions which can occur in systems having absorbing configurations $i.e.$ configurations from which the system cannot escape dynamically. The most generic universality class of APT is the directed percolation (DP) \cite{DPbook,book,DP}  class. Several model systems,  like reaction diffusion processes \cite{book}, 
depinning transitions \cite{depinning}, damage spreading \cite{spread}, 
synchronization transition \cite{synchro} and certain probabilistic cellular automata \cite{ca} 
are  known to undergo  APT belonging to  this universality class. In fact the famous `DP conjecture'  \cite{gras}  claims that any APT  with a fluctuating scalar order parameter  should generically belong 
to  DP class.  Recently  the DP  critical behaviour has  been verified  experimentally  in context of liquid crystals \cite{DPexp}.

APT in presence of a conserved field  \cite{Rossi,Heger} has drawn a lot of attention in the past decade. 
Archetypical model of an APT with a conserved density field is the conserved lattice gas  \cite{Rossi} which shows a non-DP behaviour in 1-D  \cite{Oliveira}. In general, phase transitions in systems where activity is coupled to an additional conserved field are usually believed to belong to a universality class  \cite{Rossi} different from DP. However, there are several examples of systems belonging to DP irrespective of presence of a conserved field. Of them, probably the most important is the conserved Manna model  \cite{Manna,Dickman,Lubeck}. This model, though believed to show non-DP critical behaviour for a long time, has  recently been claimed \cite{nomanna} to  belong to the DP class.  
Sticky sand-piles \cite{pk} are notable instances of DP behaviour in presence of conserved fields. In this context it is worth mentioning a different class of models, the so called threshold driven energy exchange models  which also show  \cite{cclf_ws,cclf} APTs 
different from DP where activity is coupled to a conserved energy field.

In this article we introduce a generalised energy exchange model where the energy does not respect local conservation; dissipation $\lambda$ and noise amplitude $\sigma$ control the average energy of the system.
The system shows an APT across a line of critical points $(\lambda_c,\sigma_c)$ which separates the active phase from  the absorbing  one in the $\lambda$-$\sigma$ plane. We show that the critical behaviour all along this critical line, including the special point $(\lambda=1,\sigma=0)$ where the dynamics is energy conserving, belongs to the directed percolation class.  This claim that the conserved EEM  belongs to DP,  contradicts  
some  recent studies  of related models  \cite{cclf_ws,cclf}.  We argue that the apparent non-DP behaviour  observed  earlier  is  a  consequence of  unusually long transient effects  arising due to the slow relaxation of the conserved background energy profile from random initial conditions (RIC). These  ill effects   are  avoided  here  with the help   of natural initial conditions  \cite{nomanna,natural}.  The natural initial condition is also advantageous in the non-conserved region $\lambda\simeq 1,\sigma\simeq0,$  where the long  transients associated with RIC still persist.   However, as expected, these transient baheviour  gradually disappears when dissipation  is increased.

The article is organised as follows. In the next section we define the model and study the critical point $\lambda_c$ and critical exponents at a specific noise amplitude $\sigma=1$ in the section \ref{sec:sigma1}. In section \ref{sec:phdiag} we explore the entire phase diagram in the $\lambda$-$\sigma$ plane. Section \ref{sec:cclf} is devoted to the special case of  conserved EEM, where energy density plays the role of the control parameter.

\section{The Model}\label{sec:model}
The model is defined on a periodic one dimensional lattice with $L$ sites
labelled by $i=1,2,\dots L$; each site $i$ containing  a  
positive real variable $E_i$ called energy.  A {\it pair} of neighbouring sites  is called active  when 
at least one of the sites have energy larger than or equal to  a predefined threshold $w$.
Otherwise, $i.e.$   when both these sites have energy less than $w$, the {\it pair} 
is called {\it inactive}. The  energies of an  active pair  ($i, i+1$)   
evolve following a random sequential update rule, 
\bea
E_{pair} \to   E_{pair}'&=& \lambda  E_{pair} + \xi ~;\cr
E_i  \to  r E_{pair}' &~,~&
E_{i+1}  \to  (1- r)E_{pair}',
\label{eq:noncon_dynamics}
\eea
where  the sum  of the energies  $E_{pair}= (E_i+E_{i+1})$   is  subjected to  
dissipation and  noise,  and $r \in (0,1)$  is a uniform random number.  The parameter $0\le \lambda\le1$ causes  dissipation of the total energy. With an  added  
random noise $\xi,$ chosen here  from a uniform distribution in the range $(0,\sigma),$  the system 
mimics a  dissipative particle  system  under  the stochastic force  \cite{stoch}  or  the kinetic wealth exchange models of fluctuating markets  \cite{ccm_epjb}.  Thus, like  a canonical  system  in the influence  of a Langevin bath, the  average  energy density is  expected to be controlled   by  the parameters $\lambda$ and $\sigma.$

The interesting feature of this dynamics is that it allows the possibility of an absorbing state phase transition. A configuration where  all the sites have $E_i<w,$ $i.e.$ none of the pairs are active, is an absorbing configuration of this system. Clearly, this system has infinitely many possible absorbing configurations. If the average energy density $e=\frac 1L \sum_i E_i$ is much less than  the threshold $w,$  the system is likely to fall in such a configuration.  On the other hand for very large $e$ the system remains active. The energy density $e$ is a non-decreasing function of both the dissipation $\lambda$ and the  noise amplitude $\sigma.$  Thus, for any given  $\lambda < 1,$ the average energy $e$ decreases as the amplitude $\sigma$ is decreased and one can expect an absorbing state phase transition at some critical value $\sigma_c(\lambda).$ For $\sigma > \sigma_c $ a thermodynamically large system reaches a steady state where number of active pairs in the system remains finite and for  
$\sigma \le \sigma_c $ activity certainly dies out. Alternatively, one can study this transition by keeping $\sigma$ fixed and varying the dissipation factor $\lambda$ - in that case the transition occurs at a critical value $\lambda_c(\sigma).$ 

The critical curve $\lambda_c(\sigma)$  separates the active phase from the absorbing one in the $\lambda$-$\sigma$ plane. In a later section we will study this phase diagram. In the following we first study the  critical behaviour of this system for a fixed noise amplitude, $\sigma=1,$  in some details. We shall find the critical point $\lambda_c$ and a set of  the critical exponents using Monte Carlo simulations. Since the dynamics \eqref{eq:noncon_dynamics} is invariant under the transformation $(E_i \to \mu E_i , \sigma \to \mu \sigma, w\to \mu w);$ we will work with $w=1$ without any loss of generality.

\section{Critical Behaviour for $\sigma=1$}\label{sec:sigma1}

The phase transition in this energy exchange model is characterized by the average density of active pairs    
$\rho_a = \langle \tau_i \rangle,$ where $\tau_i=1$ or $0$   depending on whether 
the {\it pair}   $(i, i+1)$ is active or inactive. In the long time limit 
$\rho_a(t)$  saturates to a steady value, which is  non zero  only  in the  
active phase; $\rho_a$ serves as the  order parameter  of this transition. 

In this section we use Monte Carlo simulations to study the absorbing transition of the EEM.
First let us expolore the critical point $\lambda_c$ and exponents for a fixed value of 
noise amplitude $\sigma =1.$

\begin{figure}[t]
 \centering
 \includegraphics[width=8.7 cm]{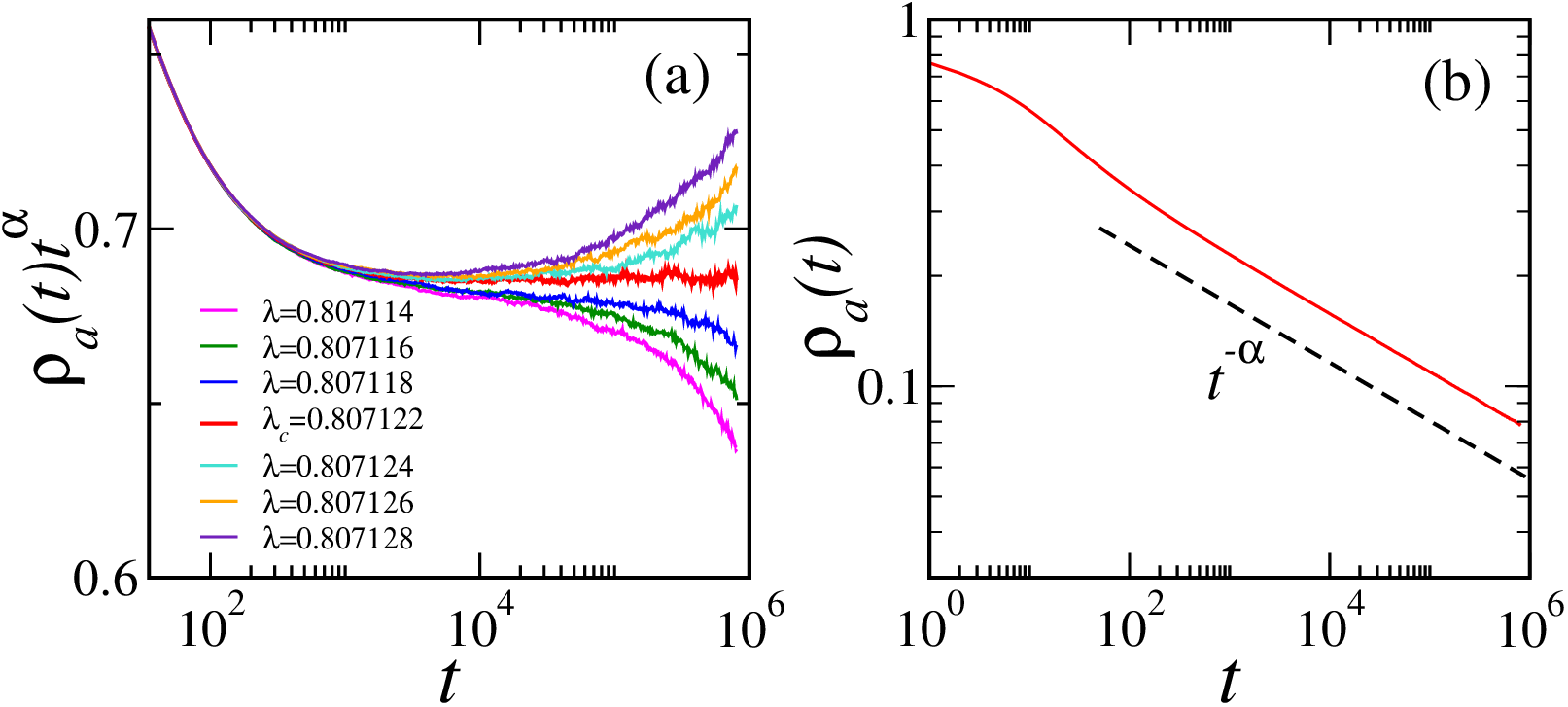}
 \caption{(Color online) Estimation of Critical point $\lambda_c$ for $\sigma=1$: (a) $\rho_a(t)t^\alpha$ versus $t$ curve for a  system of size $L=10^4$ becomes constant in the large $t$ limit for $\lambda_c=0.807122.$ (b) Estimation of $\alpha=0.159(1)$ from $\rho_a(t)$ vs. $t$ plot for $\lambda=\lambda_c.$}
 \label{fig:lc_sigma1}
\end{figure}

{\it Critical point and $\alpha$:} At the critical point $\lambda_c,$ starting from a random initial condition, the activity density decays as a power law 
\bea
\rho_a(t) \sim t^{-\alpha}.
\eea
One can estimate  the critical point $\lambda_c$  and exponent $\alpha$  
 by plotting  $\rho_a(t)$ versus $t$  for various values of 
$\lambda$ and looking for a power law decay. This estimate can be  verified 
from the plot of $\rho_a(t)t^\alpha$ against $t;$ 
the curve corresponding to $\lambda=\lambda_c$ would remain constant in the long time limit. 
This procedure is illustrated in Fig. \ref{fig:lc_sigma1} which
gives an estimate $\lambda_c=0.807122(2).$  The log scale plot of  $\rho_a(t)$ 
at $\lambda_c$ gives an accurate  estimate of the critical exponent 
\bea
\alpha= 0.159(1). \n  
\eea 
This is in very good agreement with the corresponding $\alpha_{DP}=0.15946.$

\begin{figure}[t]
\vspace*{0.6 cm}
 \centering
 \includegraphics[width=8.7 cm]{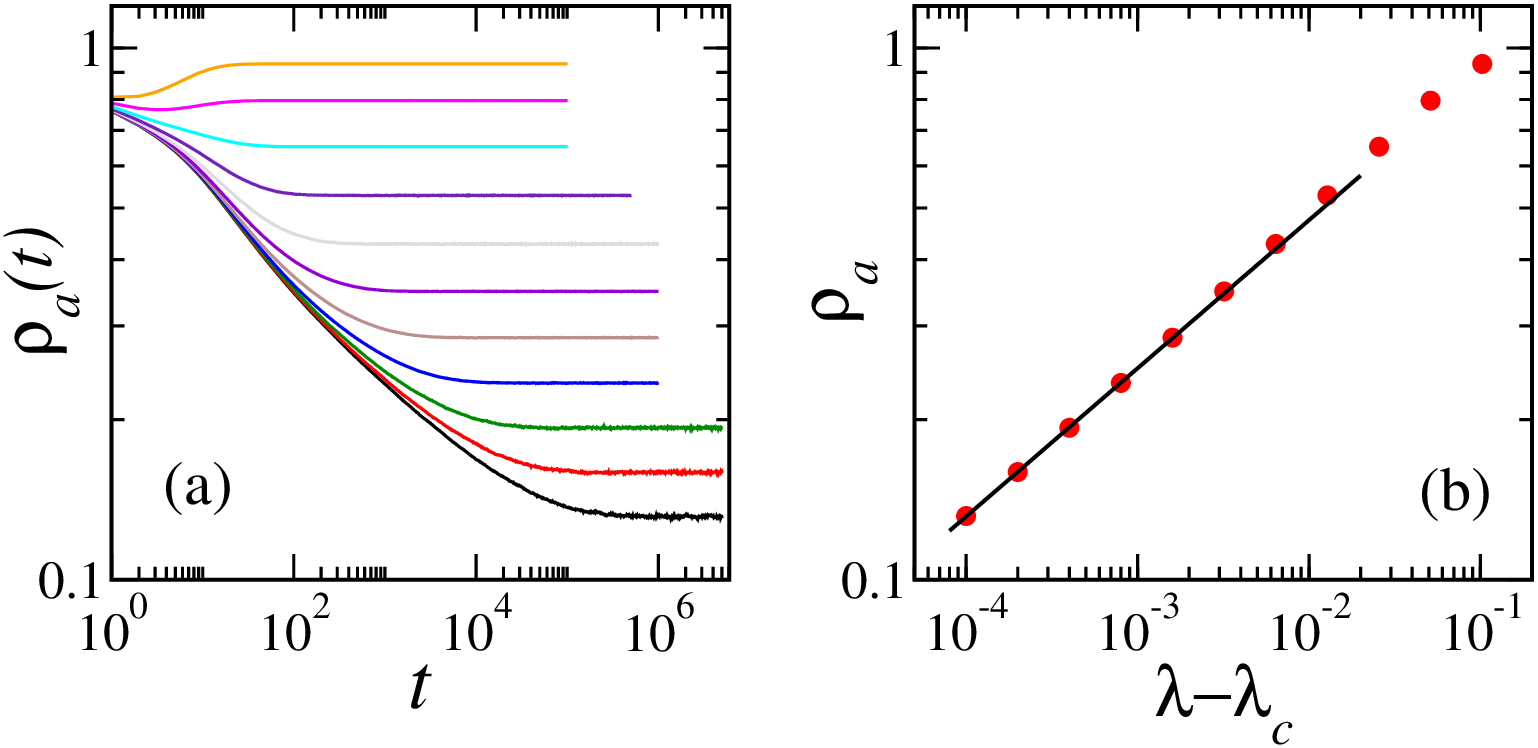}\vspace*{-0.2 cm}
 \caption{(Color online) Estimation of $\beta$ for  EEM with $\sigma=1$: (a) Plot of $\rho_a(t)$
versus $t$ for different energy densities $\lambda>\lambda_c.$ (b) Corresponding saturation values plotted 
against $\Delta = \lambda - \lambda_c$ in double logarithmic scale; the slope corresponds to $\beta=0.278.$ }
\label{fig:beta_sigma1}
\end{figure}

{\it Off-critical simulation and $\beta$:} In the active phase the activity saturates
to some finite value $\rho_a$   which vanishes algebraically  as
one approaches the critical point,
\bea
\rho_a \sim (\lambda -\lambda_c)^\beta.
\eea
Here $\beta$ is the  order parameter exponent.   
Figure  \ref{fig:beta_sigma1}(a) shows $\rho_a(t)$ as a function of $t$
for various values of $\Delta=\lambda -\lambda_c$ in the supercritical regime.  Corresponding 
saturation values  are plotted  against $\Delta$ in log-log scale (see  Fig.  \ref{fig:beta_sigma1}(b)); the slope of the resulting straight line gives us an estimate 
\bea
\beta= 0.278(2). \n 
\eea
Again, this  value of $\beta$ is consistent with  the DP value $\beta_{DP}=0.2764.$

\begin{figure}
\vspace*{0.6 cm}
\centering \includegraphics[width=8.7 cm]{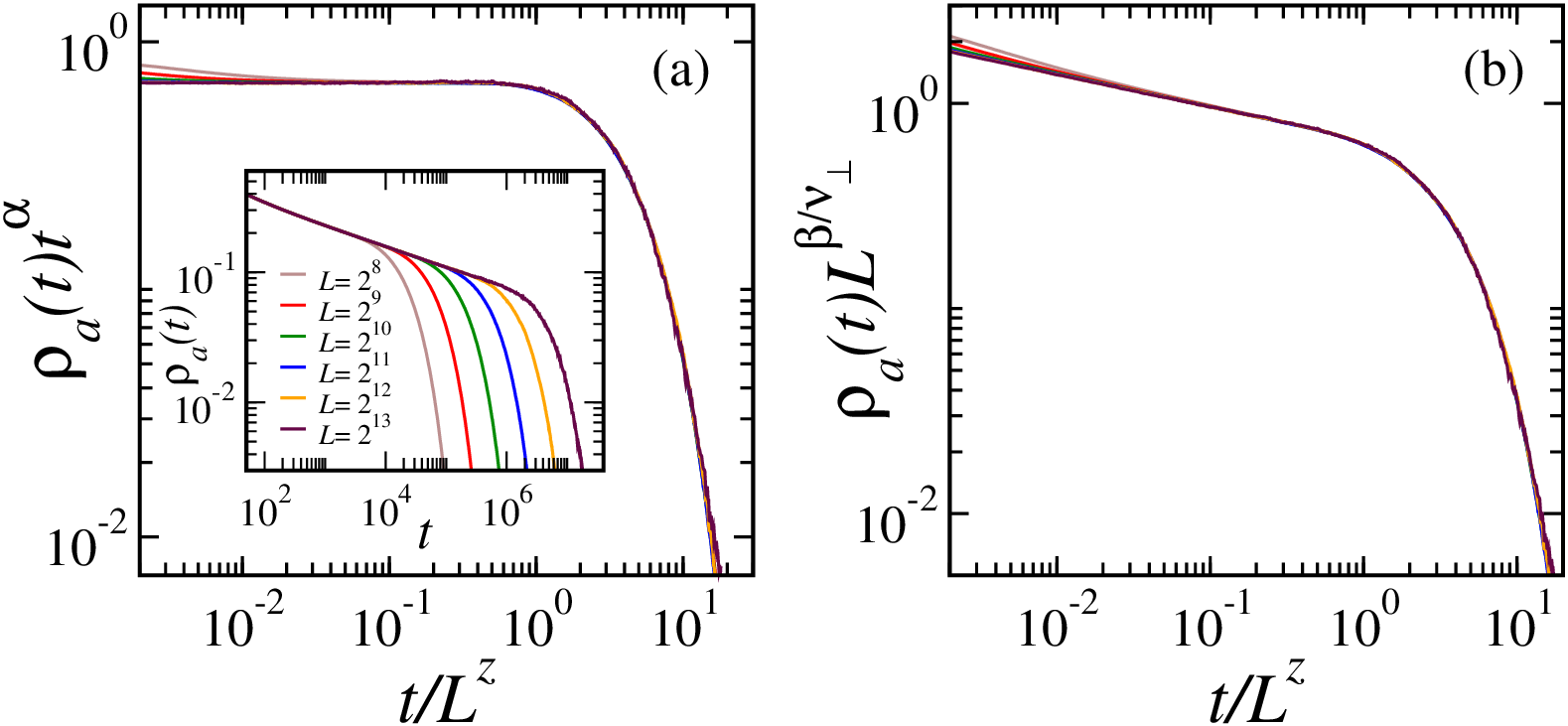}
\caption{(Color online) Finite size scaling: (a) Plot of $\rho_a(t)t^{\alpha}$ as a function of the scaled variable $t/L^z$ 
for systems of sizes $L=2^8, 2^9,\dots  2^{13}$ are  collapsed using $z=1.59(1).$ The  unscaled data are shown  in the inset. (b) The same data could be collapsed following the scaling relation \eqref{eq:z} when we use $\beta/\nu_\perp = 0.25(1).$}
\label{fig:z_sigma1}
\end{figure}

{\it Finite size scaling and $z$ :} 
Next we turn our attention to the finite size scaling. 
For a finite system, the decay of $\rho_a(t)$ at the critical point
is expected to follow the scaling forms
\bea
\rho_a(t) &=& t^{-\alpha} {\c G}(t/L^z),\label{eq:z}  \\
      &=& L^{-\beta/\nu_\perp} \tilde {\c G}(t/L^z), \label{eq:z1}
\eea
where $\c G$ and $\tilde{ \c G}$ are two different scaling functions.  
The dynamical exponent $z$ satisfies the scaling relation $\alpha z= \beta/\nu_\perp.$ At the critical point $\lambda=\lambda_c,$ both the quantities $\rho_a(t)  L^{\beta/\nu_\perp}$ and $\rho_a(t)  t^{\alpha}$
for different values of $L$ are expected to collapse on to the corresponding unique scaling curve when 
plotted against $t/L^z .$ These  data collapses, shown
in Fig. \ref{fig:z_sigma1}(a) and (b) for systems of size $L=2^8$ - $2^{13},$ 
yield 
\bea
z=1.59(1) \;\;{\rm and}\;\; \frac \beta \nu_\perp=0.25(1), \n
\eea
which are, again, in excellent agreement with the  corresponding DP  exponents.
As expected, the estimates $\alpha,$ $z$ and $\beta/\nu_\perp$ satisfy the relation $\alpha z= \beta/\nu_\perp.$

All the estimated critical exponents $\alpha, \beta, z$ and $\beta / \nu_\perp$ match quite well with the corresponding DP values suggesting that  the critical behaviour of the energy exchange model (at least for the noise amplitude $\sigma=1$) belongs to the directed percolation class. 


\subsection{The energy density}\label{subsec:energy}


Next, we  ask what happens to the  energy density $e$ when the system undergoes a phase transition. 
Will this nonorder-parameter  field $e(t)$ show  the same power-law time dependence   of  $\rho_a(t),$   as 
seen in other     models of  APT \cite{Odor} having  infinitely many absorbing states ?

\begin{figure}[t]
\vspace*{0.3 cm}
 \centering
 \includegraphics[width=8.7 cm]{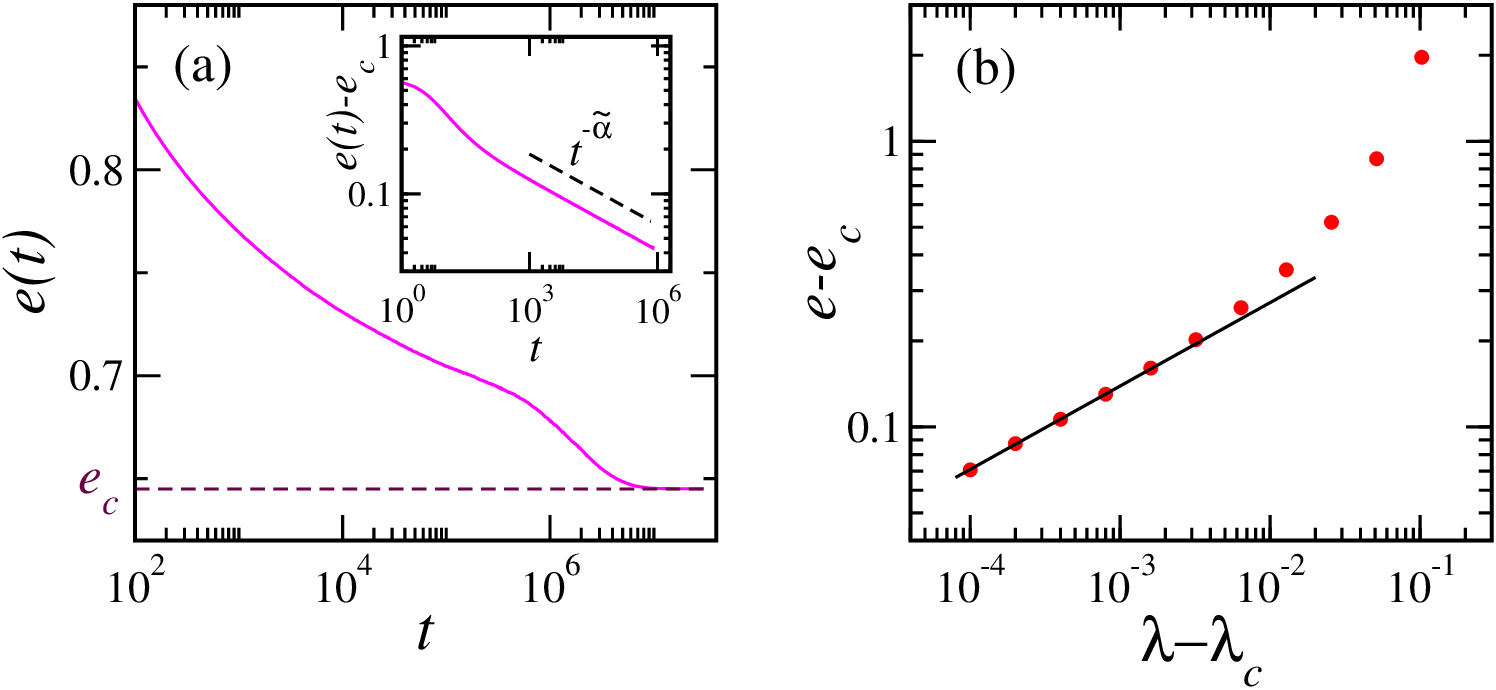}
 \caption{(Color online) (a) Plot of energy density $e(t)$ at $\sigma=1$ in the subcritical regime for different values of $\lambda=0.806322,0.806722,0.806922,0.807022$ and $\lambda_c.$
  $e(t)$ saturates to $e_c=0.645$ at  and below the critical point $(\lambda_c=0.807122,\sigma_c=1).$ The inset shows the corresponding log scale plot of $e(t)-e_c;$ the slope, measured over last three decades, results in an estimate $\tilde \alpha=0.158(1).$ (b) The steady state values of the reduced energy $e-e_c,$ plotted against $\lambda -\lambda_c$ for $\sigma_c=1,$ estimates $\tilde \beta = 0.292(3).$ }
 \label{fig:energy}
\end{figure}

In the  active phase the energy density $e(t)$  evolves along with  $\rho_a(t),$  and saturates to a stationary value $e(\lambda,\sigma)$ in the long time limit. The critical value $e_c \equiv e(\lambda_c,\sigma_c)$ can be obtained from the decay of $e(t)$ as shown in  Fig. \ref{fig:energy}(a) for $\lambda_c=0.807122$ and $\sigma_c=1.$ As $t \to \infty$ $e(t)$ saturates to a value $e_c =0.6452(4).$ 
We find that the reduced energy density $e(t)-e_c$ shows an algebraic decay $t^{-\tilde \alpha};$  this is illustrated in the inset of Fig. \ref{fig:energy}(a). A linear fit near the critical point yields 
\bea
\tilde \alpha = 0.158(1). 
\eea

 Clearly, within the error bars $\tilde \alpha$ is not different from $\alpha_{DP}= 0.1594.$ 
This indicates that the reduced energy $e-e_c$ may be considered as an alternative order parameter of the APT in the EEM. To check whether the critical exponents of energy density indeed belongs to the DP class, we have also estimated the order parameter exponent $\tilde \beta.$ A plot of $e(\lambda)-e_c(\lambda_c)$ against $\lambda - \lambda_c$ in log scale for $\sigma=1$ is shown in Fig \ref{fig:energy}(b); the slope of the straight line gives

\bea
\tilde \beta= 0.292(3).
\eea
Once again, we find that $\tilde \beta \simeq \beta_{DP}.$ The other exponents like $z$ and $\beta/\nu_\perp$ for density  are also found to be consistent with the corresponding DP values (data not shown here).  Thus, the APT in EEM can also be characterized by a non-order parameter $e-e_c$ similar to some other models with infinitely many absorbing configurations \cite{Odor,Mohanty}.

\section{Phase Diagram}\label{sec:phdiag}

\begin{figure}[t]
\vspace*{0.6 cm}
 \centering
 \includegraphics[width=8.5 cm]{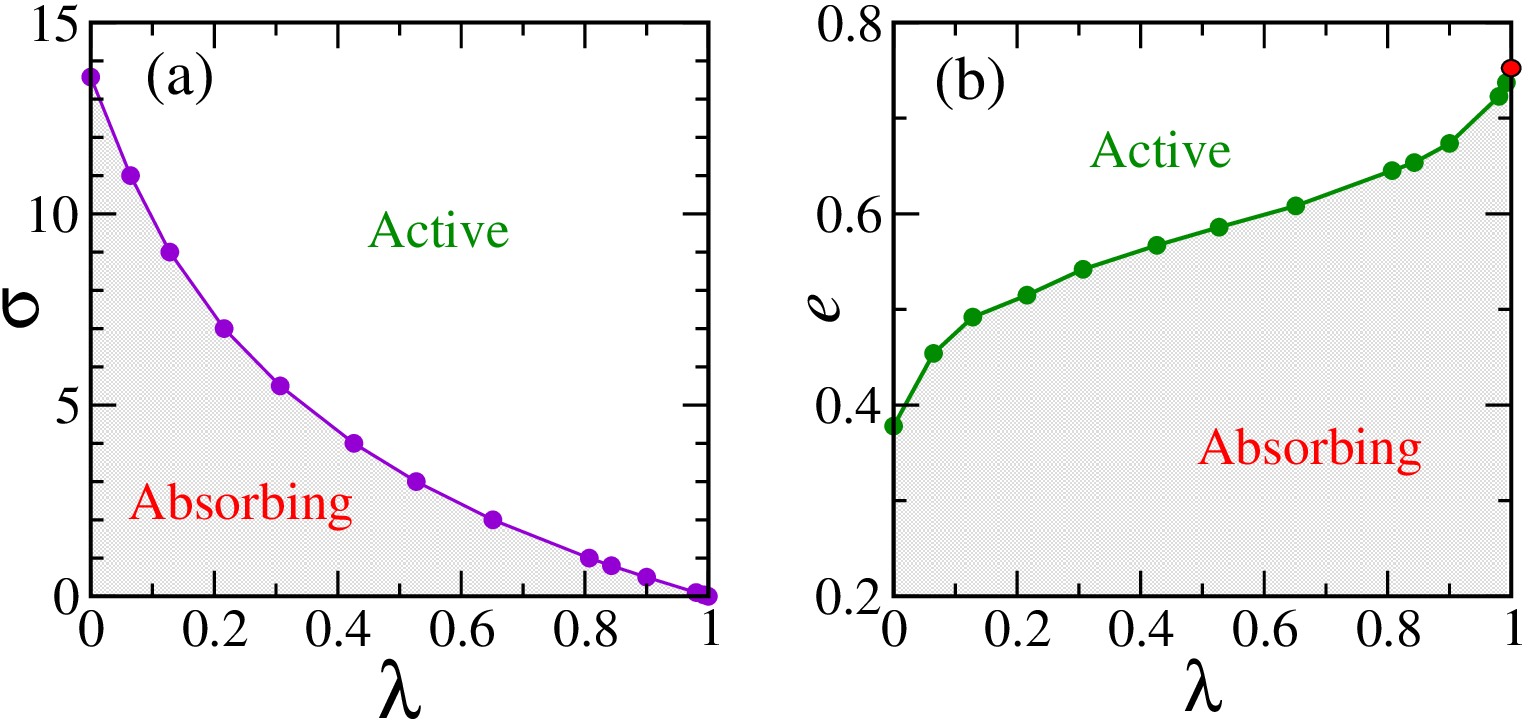}
 \caption{(Color online) (a) Phase diagram of EEM in the $\lambda$-$\sigma$ plane. The critical line  separates the absorbing phase (shaded region) from the active one. (b) Corresponding phase diagram in the $\lambda$-$e$ plane. The point marked red corresponds to the critical energy of the conserved model.}
 \label{fig:phase}
\end{figure}

The average energy of EEM is controlled by two parameters $\lambda$ and $\sigma.$  This gives rise to a phase diagram in the two dimensional $\lambda$-$\sigma$ plane where the active and inactive phases are separated by the line of critical points $(\lambda_c, \sigma_c).$ We have used Monte Carlo simulations to trace this critical line in the phase plane. For a set of values of the noise amplitude $\sigma,$ we have estimated the critical point $\lambda_c$ using the procedure described in the previous section. These critical values $(\lambda_c, \sigma_c)$  are listed in the Table \ref{tab:phdiag}. 
The critical line, obtained by joining these points in the phase plane,is plotted in Fig \ref{fig:phase}(a).
The shaded region $ \sigma < \sigma_c(\lambda)$ corresponds to the absorbing phase of the system.

The phase diagram can also be drawn in the $e$-$\lambda$ plane eventhough $e$ is not an external tuning parameter. The steady state value of energy at the critical point $e(\lambda_c,\sigma_c)$ can be considered as the critical point $e_c$ for a given $\lambda=\lambda_c.$ The values of  $e_c$ are also listed in Table \ref{tab:phdiag} along with $(\lambda_c, \sigma_c).$ Note that the statistical errors in the estimates of $e_c$
are comparatively large. Figure \ref{fig:phase}(b) shows the phase diagram in the $e$-$\lambda$ plane.


\begin{table}[t]
\begin{center}
 \begin{tabular}{|c|c|c|}
\hline
$\sigma_c$ & $\lambda_c$ & $e_c\equiv e(\lambda_c,\sigma_c)$ \cr
\hline
0.04 & 0.992117(1)  & 0.7372(2)\cr
0.1  & 0.980089(1)  & 0.7228(2)\cr
0.5  & 0.90000(2)   & 0.6736(4)\cr
0.8  & 0.84305(1)   & 0.6536(4)\cr
1  & 0.807122(2)  & 0.6452(4)\cr
2  & 0.65113(1)   & 0.6089(4) \cr
3  & 0.526988(4)  & 0.588(4)\cr
4  & 0.42618(2)   & 0.569(2)\cr
5.5  & 0.30667(1)   & 0.542(1)\cr
7  & 0.21592(2)   & 0.515(3)\cr
9  & 0.12808(2)   & 0.490(2)\cr
11 & 0.06450(2)   & 0.454(1)\cr 
13.578(2) & 0       & 0.378(1)\cr
\hline
0    & 1            & 0.75243(3) \cr
\hline
 \end{tabular}
\end{center}
\caption{Critical points of the $(1+1)$-dimensional energy exchange model. $e_c$ gives the average energy density at the critical point. The last row corresponds to the conserved case. }\label{tab:phdiag}
\end{table}

We have studied   the decay of $\rho_a(t)$   for  all these points (listed in Table \ref{tab:phdiag})   
and  find that   the critical behaviour of EEM is   consistent with DP-universality on 
the entire critical line.  We have not reported  these results  in details here as they 
are only repetitions of the same excercise done  in the previous section for $\sigma=1$.

Note that the critical line approaches the point $(\lambda=1,\sigma=0)$ as the noise amplitude $\sigma$ is decreased. This is expected as for $\sigma=0,$ the system always falls into an absorbing state for any $\lambda<1$ by continuously dissipating energy. In this case, when $\lambda=1,$ the energy is not disspated from the system and thus  the dynamics becomes energy conserving; the total energy is fixed by the initial condition. To study the phase transition at this special point $(\lambda=1,\sigma=0)$ one must tune the conserved energy density $e.$  For  the conserved system the background energy  profile does not evolve rapidly. The fluctuations  existing in random initial  configurations  persists for a  long  time  and  
the relaxation of the system  to the stationary state, where the energy profile is essentially  flat,
becomes very slow. A separate section is devoted for a careful study of this conserved EEM.

\begin{figure}[t]
\vspace*{0.5 cm}
 \centering
 \includegraphics[width=8.7 cm]{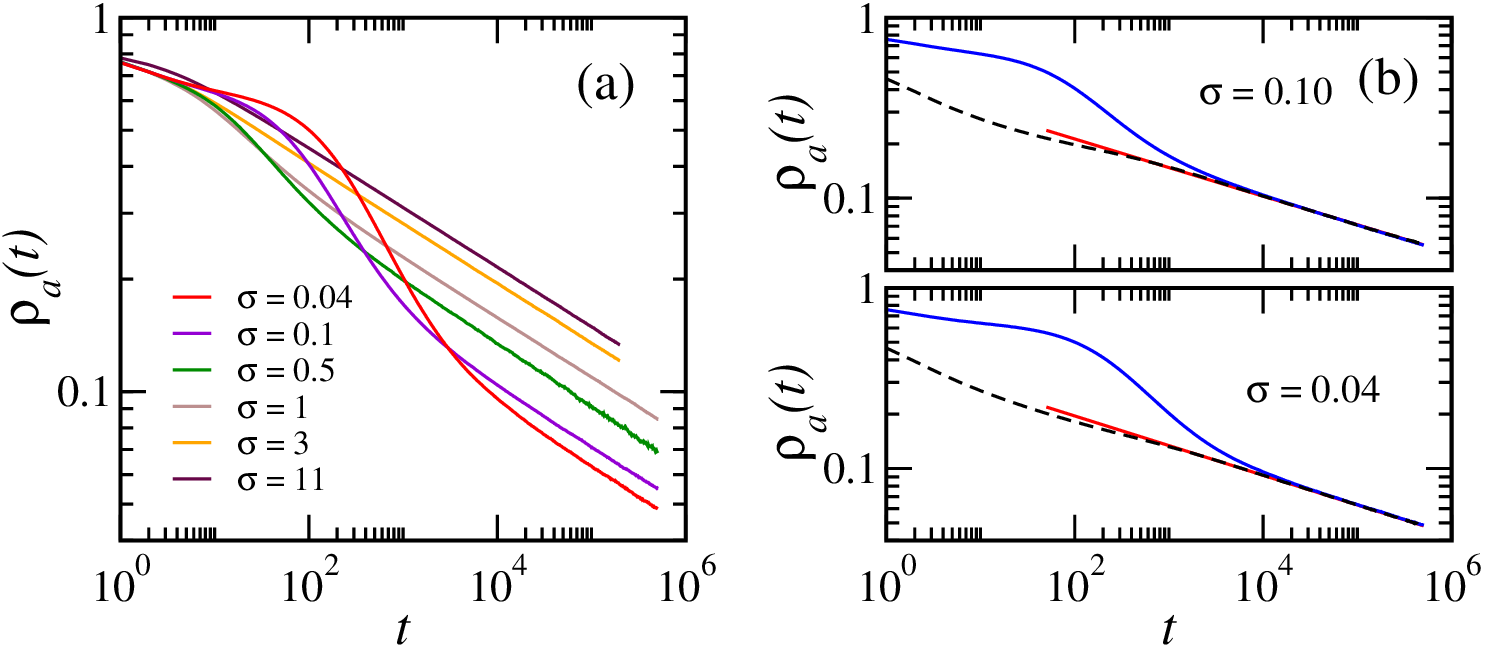}
 \caption{(Color online) (a) Decay of activity $\rho_a(t)$ at the critical point for random initial condition for different noise amplitude $\sigma.$ The power law behaviour starts at late times as $\sigma \to 0$ (or $\lambda \to 1$). (b) Comparison of $\rho_a(t)$ starting from random (blue solid line) and natural  (black dashed line) initial conditions for two different  noise amplitudes  $\sigma = 0.04$ (lower panel) and $\sigma= 0.10$ (upper panel). The solid red line corresponds to slope $\alpha.$ The system size is $L=10^4.$}
 \label{fig:nat}
\end{figure}

These  transient effects are  also  pronounced in the vicinity of the special point $(\lambda=1,\sigma=0).$  
For example, $\rho_a(t)$  shows an atypical decay from  random initial conditions.  
 In Fig. \ref{fig:nat}(a) we have plotted $\rho_a(t)$ versus $t$ curve for a set of points $(\lambda_c,\sigma_c)$ on the critical line.  Evidently,  the algebraic decay $\rho_a(t) \sim t^{-\alpha}$ starts at an increasingly longer timescale as $\lambda_c$ 
  is increased.  In particular, substantial numerical effort is required to study the  critical behaviour near the conserved limit  $\lambda_c=1$  which is reasonably reduced  if one uses the so called {\it natural initial conditions} \cite{nomanna,natural}.

 Natural initial conditions are  prepared by  reactivating   the  steady state  configurations. Thus, they have   the same correlations .
existing in the stationary state.
For any value of the control parameter $(\lambda, \sigma),$ one  usually starts from a random initial configuration and let the system evolve to the steady state. The natural initial states are then prepared from these steady state configurations by allowing the system to diffuse for a short time interval during which the energy of any randomly selected site, independent of  whether it is active or not,  is  distributed unbiasedly among its neighbours. This process creates enough activity in the system so that one can observe the decay, but does not destroy the natural correlations built in the stationary state. 
Figure \ref{fig:nat}(b) shows a comparison of decay of activity starting from random (solid) and natural (dashed) initial conditions for two small values of $\sigma=0.04$ and $0.1$ for which this transient effect is most pronounced.  Clearly, the curves corresponding to the natural initial condition reaches the scaling regime  earlier than the same for random ones.

 It is natural to expect that the conserved model ($\lambda=1, \sigma=0 $)  suffers   worst from these ill-effects
of random  initial conditions.  The need to use natural initial conditions will become more apparent in the next section where  
we turn our attention to this conserved energy exchange model and study APT by varying the conserved energy density $e.$

\section{The Conserved Energy Exchange Model}\label{sec:cclf}

The dynamics of the EEM is energy conserving at the special point $(\lambda=1,\sigma=0)$  $i.e.$ when energy is neither dissipated nor added to the system as noise. In this case, the active pairs of neighbouring sites reshare their energies following,
\bea
E_i & \to & r(E_i+E_{i+1}) \cr
E_{i+1} & \to & (1- r)(E_i+E_{i+1})
\label{eq:conserved_dynamics}
\eea
where $r \in (0,1)$ is a uniform random number.
Thus, the total energy does not evolve in time. This conserving dynamics have been studied in absence of a threshold ($i.e.$ when $w=0$) in different contexts of heat transport  \cite{kipnis} and Econo-physics  \cite{yakovenko} earlier. Like the non-conserved model, this conserved EEM also undergoes an APT but the control parameter here is the conserved energy density $e=\frac 1L \sum_i E_i.$ Here too the density of active pairs $\rho_a$ plays the role of the order parameter and attains a non-zero stationary value only beyond some critical energy density $e_c.$

It has long been argued that absorbing transitions in presence of conserved fields  \cite{Heger,Lubeck} belong to a universality class different from DP  \cite{Rossi}. But there are examples where an absorbing transition depicts DP behaviour even in presence of additional conserved fields  \cite{pk,nomanna}. In this view, it is interesting to explore the critical behaviour of the special case $(\lambda=1, \sigma=0)$ of the energy exchange model.

\begin{figure}[t]
\vspace*{0.2 cm}
 \centering
 \includegraphics[width=8.5 cm]{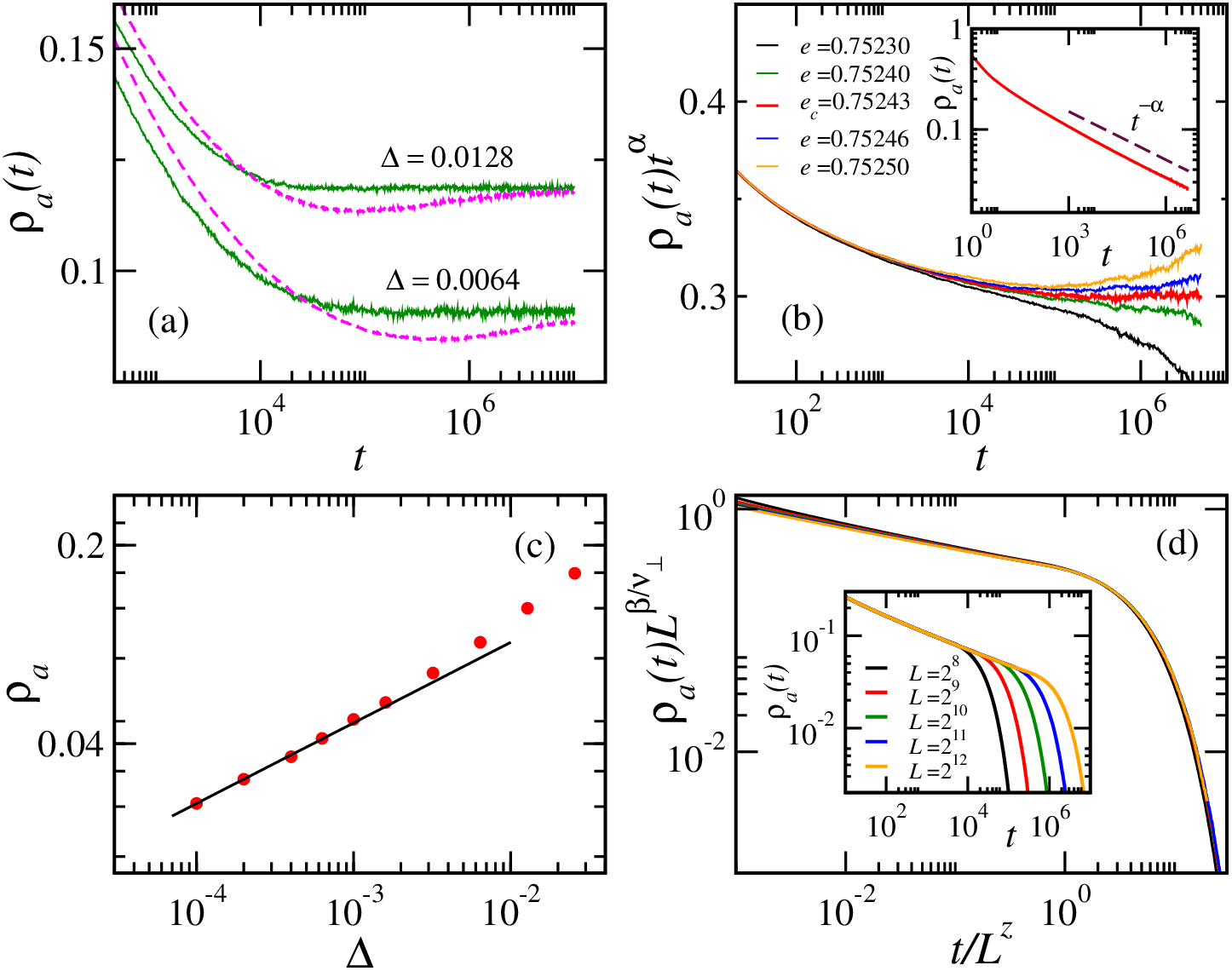}
 \caption{(Color online) Conserved EEM (a) Undershooting: Comparison of decay of $\rho_a(t)$ starting from random (dashed magenta) and natural (solid green) initial conditions for a system of size $L=10^5.$ (b) Determination of $e_c$:  Plot of $\rho_a(t)t^\alpha$ versus  $t$ for different values of $e;$ the horizontal curve corresponds to the critical point $e_c=0.75243.$  The inset shows log scale plot of $\rho_a(t)$ versus $t$ for $e=e_c$ with slope  $\alpha=0.159.$ (c) Estimation of $\beta$: The saturation values of activity $\rho_a$ are plotted against $\Delta = e-e_c$ in double logarithmic scale; the slope corresponds to $\beta=0.283.$ (d) Finite size scaling: $\rho_a(t)L^{\beta/ \nu_\perp}$ as a function of $t/L^z$ for different $L$ are collapsed  using $z=1.58$ and $\beta/ \nu_\perp = 0.26.$ The inset shows the unscaled data.}
 \label{fig:cclf}
\end{figure}

First let us check whether $\rho_a(t)$ shows any unusual transient behaviour.
We have measured the decay of activity $\rho_a(t)$ starting from random initial conditions, where the total energy is distributed randomly among all the lattice sites,  for two different values of energy density $e$ which are shown as dashed lines in Fig. \ref{fig:cclf}(a). The pronounced undershooting seen in these curves is an artifact of the disordered random initial conditions. As discussed in the previous section, one can avoid this long transient effect if `natural' initial conditions are used. These initial conditions, prepared by taking the system to a stationary state and then allowing the energy to diffuse for a short time, lead to well behaved decay profile $\rho_a(t)$ (see solid lines in Fig. \ref{fig:cclf}(a)). 
Since it is known  that presence of undershooting  in  $\rho_a(t)$ may lead to erroneous estimation  of the critical point and subsequent determination of critical exponents  \cite{nomanna}, in the following we study the APT of the conserved EEM using natural initial conditions.

\subsection*{Critical behaviour and exponents}

First, let us estimate the critical point $e_c$ following the same procedure 
discussed in section  \ref{sec:sigma1}. The plot of $\rho_a(t) t^\alpha$  versus $t$ (shown for different   
$e$ in Fig. \ref{fig:cclf}(b)) saturates in the long time limit for $e_c=0.75243(3).$  The  $\rho_a(t)$ versus $t$ plot at this critical energy $e_c$ gives an estimate of the decay exponent 
\bea
\alpha= 0.159(2). \n 
\eea 
This value of $\alpha$ is consistent with $\alpha_{DP}.$


The order parameter exponent $\beta$ is determined from the steady state values $\rho_a$ of the activity 
in the active phase, as $\rho_a \sim (e-e_c)^\beta.$ Figure  \ref{fig:cclf}(c) shows a plot of $\rho_a$ versus $\Delta=e-e_c$  in log-log scale; the slope of the resulting straight line gives us an estimate 
\bea
\beta= 0.283(4).\n
\eea
Again this  value of $\beta$  is 
consistent with  the DP value $\beta_{DP}=0.2764.$  

The dynamical exponent $z$ and $\beta \nu_\perp$ are determined using the scaling form \eqref{eq:z}
by plotting $\rho_a(t)  L^{\beta/\nu_\perp}$ against $t/L^z$ for different values of $L$  and looking for a data collapse. The best collapse is shown in Fig. \ref{fig:cclf}(d) for systems of size $L=2^8$-$2^{12},$ and 
yields  an estimate  
\bea
z=1.58(2) ~~;~~~  \frac \beta \nu_\perp= 0.26(1) \n
\eea
which are, again, in excellent agreement with the  corresponding DP  exponents.

We find that all the critical exponents $\alpha,\beta, z$ and $\beta/\nu_\perp$ 
for the conserved energy exchange model agree quite well with the 
corresponding DP values.
Thus  this  system provides another example   where the presence of an additional conserved field does not a induce a different critical behaviour. It may be mentioned  that   the  fixed energy sandpiles \cite{nomanna} 
are  known to be in  DP class albeit having a conservation.

\subsection{The Minimal Model}

A variant of the energy exchange model has also been studied  \cite{cclf_ws} recently,
where a pair of neighbouring sites randomly re-share their energies following the energy conserving dynamics \eqref{eq:conserved_dynamics} when at least one of them has energy \textit{less} than or equal to the threshold $w.$ This model, henceforth referred to as the minimal energy exchange model,  also undergoes an  absorbing phase transition,  apparently  showing a non-DP behaviour  \cite{cclf_ws}. Since  the minimal model has the identical local dynamics \eqref{eq:conserved_dynamics}, it is natural to expect that its critical behaviour  is  same as the conserved EEM  studied in Sec - \ref{sec:cclf}. The non-DP  behaviour  observed  for the 
minimal model could have resulted from the long transients present in random initial conditions. In  view of this we  briefly revisit this  model  and  study  the critical behaviour using natural initial conditions.

\begin{figure}[ht]
 \centering
 \includegraphics[width=8.5 cm]{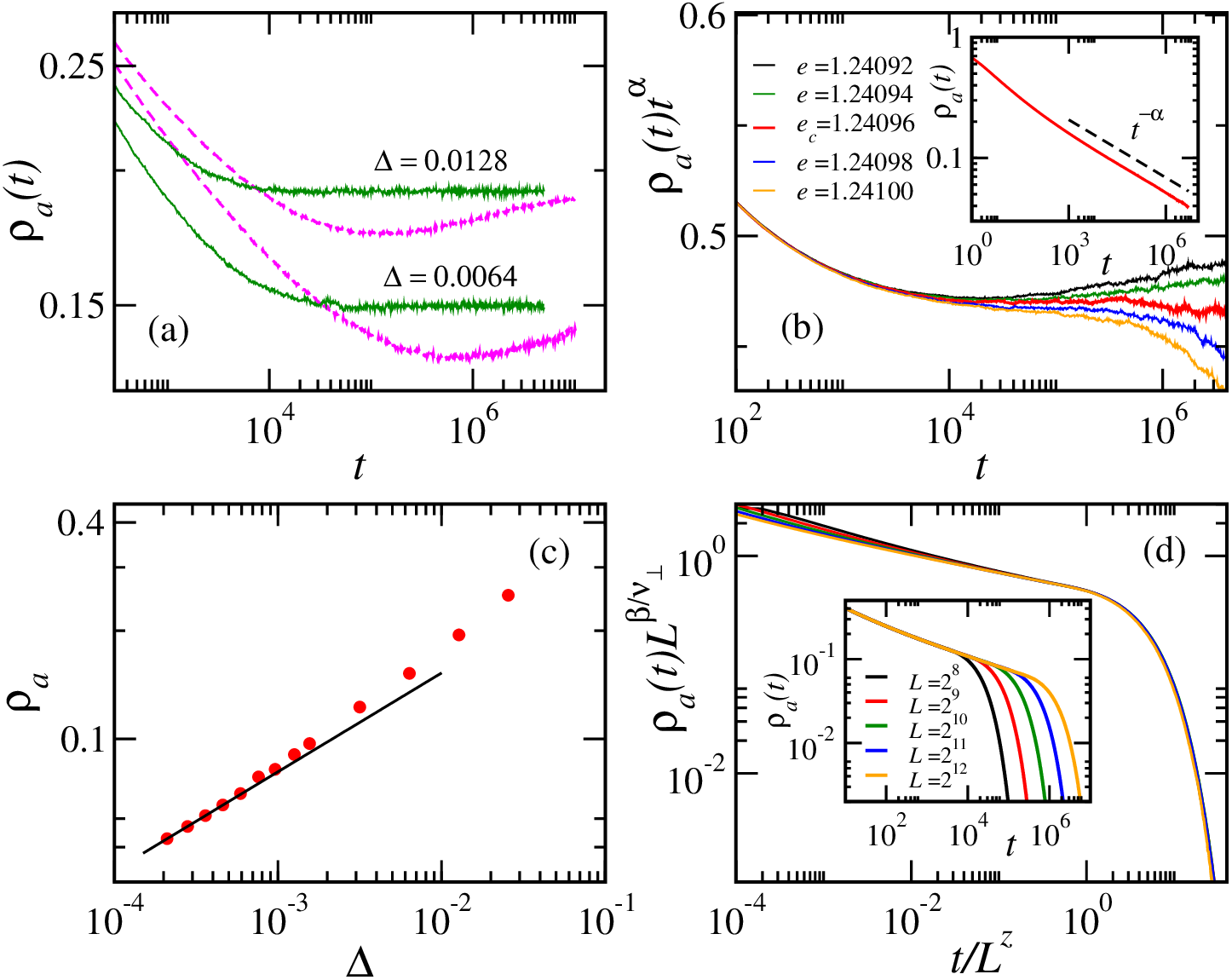}
 \caption{(Color online) Minimal conserved EEM (a) Undershooting: Comparison of $\rho_a(t)$ versus $t$ plot starting from random (dashed magenta) and natural (solid green) initial conditions; system size $L=10^5.$ (b) Critical point $e_c$: $\rho_a(t)t^\alpha$ versus $t$ curve for a  system of 
size $L=10^5$ becomes constant for $e_c=1.24096$ as $t \to \infty.$  The inset shows estimation of $\alpha=0.159$ from $\rho_a(t)$ vs. $t$ plot. (c) Estimation of $\beta$: Saturation values of activity $\rho_a$ plotted as a  function of $\Delta=e_c-e;$ the slope of the solid line corresponds to $\beta=0.275.$ (d) Finite size scaling: Data collapse according to Eq. \eqref{eq:z} could be obtained when $z=1.54$ and $\beta/ \nu_\perp = 0.25.$ The inset here shows the unscaled data.} \label{fig:cclf_min}
\end{figure}

For completeness  let us   define  the minimal model  explicitly in one dimension.  On a periodic  lattice  
of size $L,$  each site  $i$     has energy $E_i.$  
A    pair  of neighbouring sites $(i,i+1)$ is  said 
to be active  when  at least one of the sites $i$ or $i+1$ has energy less than or equal 
to  a  threshold value $w$, set to be unity. The active pairs in the model 
evolve following the energy conserving random sequential dynamics \eqref{eq:conserved_dynamics}.

We proceed by defining the density of active pairs  $\rho_a$ as the  
order parameter.   The decay of activity $\rho_a(t)$   in the  super critical  regime, starting 
from the natural initial condition (solid line),  is shown  in Fig. \ref{fig:cclf_min}(a)  for  two different values of average energy $e.$ Clearly  $\rho_a(t)$  is well behaved  and  approaches to a  
stationary value reasonably  fast.  For comparison, in the same figure,  we have  included  
plot of  $\rho_a(t)$  from  the random initial condition (dashed line) for the same values of $e.$
Evidently, the  pronounced undershooting  in   the random initial condition  becomes  stronger as one 
approaches  the critical point. These effects may lead to inaccurate determination of 
$\alpha $ and $e_c$.  Here we  study the critical behaviour  using natural initial condition.

Figure \ref{fig:cclf_min}(b) illustrates the scheme for determination of the critical point which yields 
\bea
e_c=1.24096(3). \n
\eea  
This  estimate  of critical point $e_c$  is to be compared with 
$1/w_c$  reported in  \cite{cclf_ws}  for the following reason. 
The minimal model  was studied earlier  with a fixed $e=1$  and varying  
the threshold $w.$ This resulted in a critical point $w_c=0.810$; 
in other words  in the $(e,w)$ phase plane $(1, w_c)$ is a critical point. 
It is evident that    $(\lambda , \lambda w_c)$  is also a critical point 
for any arbitrary  $\lambda$ as both the  conditional statement $E_i<w$ and 
the dynamics  \eqref{eq:conserved_dynamics} are invariant  under a scale transformation
$(E_i \to \lambda E_i ,  w\to \lambda w).$ This implies  that 
the   critical line  in the $( e,  w)$ phase plane is  a straight 
line with unit slope  passing through  $(1, w_c)$.  Clearly $(1/w_c, 1)$ is also a 
critical point and it can be  reached   keeping  $w$ fixed; thus of the minimal model along the line $w=1$
is  expected to  show the transition at $e_c=1/w_c$  when $e$ is  tuned.

 The critical  point $e_c=1.24096(3)$ is slightly  higher than  the previously 
reported  value $1/w_c =1.23456$  \cite{cclf_ws}. We further show that this nominal correction in the  estimate of critical point  results in   revised critical exponents  which  agree remarkably  well   
with DP  universality class.

The slope of  the  curve  $\rho_a(t)$  versus $t$ in log scale 
(over last two decades) at the critical point $e_c$, shown in the inset of Fig. \ref{fig:cclf_min}(b),  
gives an estimate of the exponent 
\bea
\alpha= 0.159(2),\n 
\eea
which is in strikingly good agreement  with $\alpha_{DP}.$

The stationary densities  $\rho_a$  
plotted against $e-e_c$ in log scale in Fig \ref{fig:cclf_min}(c) 
results in  an  estimate  
\bea
\beta = 0.275(1), \n 
\eea
which, again, matches well with $\beta_{DP}.$

The   dynamical exponent $z$  and $\beta/\nu_\perp$ are obtained using the standard finite size scaling collapse following Eq. \eqref{eq:z}. Figure \ref{fig:cclf_min}(d) shows this data collapse for  $L=2^8 - 2^{12}$ using  
 \bea
 z= 1.54(2) ~~~{\rm and}~~~ \frac \beta \nu_\perp = 0.25(1). \n
 \eea
Both the estimates are in good agreement with the corresponding DP 
values.

To summarize,  the critical exponents of the minimal model 
imply that, contrary to previous claims  \cite{cclf_ws}, 
the APT seen here, in fact, belongs to DP class.  We  must mention that the differences in the estimates of critical exponents found in this study are  not in anyway related to the fact that the order parameter in 
Ref.  \cite{cclf_ws}  was  chosen differently. Instead  of  density 
of {\it active  pairs} $\rho_a$,  the average density  of
``sites having $E_i\le w$'' was used  as order parameter.  This order parameter  
  $\rho=  \langle s_i\rangle$, where $s_i=0$ when $E_i> w$ or otherwise $s_i=1,$
is related to $\rho_a$ as 
\bea 
\rho_a=  \rho +   \langle s_i (1-s_{i+1})\rangle.
\eea
Since  $\rho_a$ is non-zero  only when $\rho \ne 0$ and 
they are  dimensionally identical,  the  critical point  and  
corresponding exponents  would be same in both cases.

\section{Summary}\label{sec:summary}

In conclusion, we have studied absorbing phase transition  in 
the energy exchange models in  one spatial dimension. The dynamics of this model 
is controlled by two parameters - a  noise of amplitude $\sigma$ and 
a dissipation factor $\lambda.$ The absorbing and active phases are separated by a critical line in the $\lambda$-$\sigma$ plane. With extensive Monte Carlo simulations we show that 
the critical behaviour of the energy exchange model belongs to directed percolation along the entire critical line. The critical line    ends at the  point  $(\lambda=1, \sigma=0)$, where the   dynamics is energy conserving. The numerical study  of  the critical behaviour suffers  from   long transients and unusual decay profiles when this conserved limit is  approached. 
We show that this effect can be removed if one uses suitably prepared natural initial conditions. 

The conserved EEM $(\lambda=1, \sigma=0)$  is special in a way that 
the  conserved energy itself  serves  as the  tuning parameter.   
This conserved version  also suffers from  the presence of long transients 
 making it numerically hard to study the system using random initial conditions. Here again one can take advantage of the natural initial conditions  which reaches the stationary state within a reasonably shorter time. We find that the critical behaviour of the conserved EEM also belongs to DP.

We also revisit the minimal energy exchange model (a variant of conserved EEM),  
and find that  this model too shows DP critical behaviour. The apparent non-DP exponents found 
earlier  \cite{cclf_ws}  are possibly due to the undershooting present in 
the decay of activity from the random initial conditions.

With this study we  want to emphasize the fact that one has to be very 
careful in exploring absorbing phase transition in presence of a 
conserved field. The  conserved field  imposes additional constraints   
on  the evolution of the system and  introduces a long time scale. 
Random initial conditions   may need very long time
to saturate, producing  unwanted  transient effects 
like undershooting. These undesirable  features  may lead to erroneous  
estimation of critical point and  hence inaccurate critical exponents. 
One must take care of these ill effects  while studying APT  in  a  
system  with a  conserved field. 

{\bf Acknowledgement:} U.B. would like to acknowledge thankfully the financial support of the
Council of Scientific and Industrial Research, India. (Grant No. SPM-07/489(0034)/2007).

\end{document}